\documentclass[aps,prb,groupedaddress,superscriptaddress,twocolumn]{revtex4}

\usepackage{graphicx, amsmath, subfigure}
\usepackage{amsfonts}
\usepackage{hyperref}
\usepackage{color}
\usepackage{float}
\usepackage{chngcntr}
\usepackage{mathrsfs}

\begin{document}

\preprint{}


\title{Non-Abelian operation through scattering between chiral Dirac edge modes}


\author{Zhi-Xing Lin}
\affiliation{International Center for Quantum Materials, School of Physics, Peking University, Beijing 100871, China}

\author{Yijia Wu}
\thanks{Corresponding author: yijiawu@pku.edu.cn}
\affiliation{International Center for Quantum Materials, School of Physics, Peking University, Beijing 100871, China}

\author{X. C. Xie}
\affiliation{International Center for Quantum Materials, School of Physics, Peking University, Beijing 100871, China}
\affiliation{CAS Center for Excellence in Topological Quantum Computation, University of Chinese Academy of Sciences, Beijing 100190, China}

\date{\today}

\begin{abstract}

We theoretically demonstrate that non-Abelian braiding operation can be realized through the scattering between chiral Dirac edge modes (CDEMs) in quantum anomalous Hall insulators by analytically deriving its $S$-matrix. Based on the analytical model, we propose a viable device for the experimental realization and detection of the non-Abelian braiding operations. Through investigating the tunneling conductance in a discretized lattice model, the non-Abelian properties of CDEMs could also be verified in a numerical way. Our proposal for the CDEM-based braiding provides a new avenue for realizing topologically protected quantum gates. 

\end{abstract}

\maketitle



\section{Introduction}

Majorana zero modes (MZMs) are self-conjugate excitations which appear as vortex-bound states\cite{Ivanov} or topologically protected end states\cite{kitaev2001unpaired,alicea2011non} in topological superconductors. MZMs are regarded to possess non-Abelian statistics\cite{alicea2011non} due to its topology-related non-Abelian geometric phase accumulated during the braiding operations. Owing to these charming properties, MZMs have been widely investigated\cite{wang2018evidence,zhu2020nearly,kong2021emergent, Delft2012signatures,rokhinson2012Majorana,deng2012anomalous,churchill2013superconductor,das2012zero,finck2013anomalous} and regarded as the most promising candidate for topological quantum computation\cite{sarma2015majorana}.

The experimental signals for MZMs have been experimentally reported in various platforms such as the iron-based superconductors with vortices\cite{wang2018evidence, zhu2020nearly, kong2021emergent} and the semiconductor nanowires proximate to $s$-wave superconductor\cite{Delft2012signatures,rokhinson2012Majorana,deng2012anomalous,churchill2013superconductor,das2012zero,finck2013anomalous}. Since MZMs are ``zero-dimensional'' localized states bound to the vortex or the end of the nanowire, the braiding operation of MZMs are expected to be conducted through spatially moving the vortices (in the assistance of the probing technology such as scanning tunneling microscope tips) or modulating the gate voltages in superconductor-semiconductor nanowire junctions\cite{alicea2011non, sau2011controlling, van2012coulomb}. However, the adiabatic condition\cite{PhysRevB.91.174305} will impose restriction on the braiding operation that the typical time cost for each braiding operation should be in the scale of nanosecond\cite{PhysRevB.91.174305}. Moreover, extra instability could also be introduced since the braiding operations are conducted through external manipulations, in which some undesirable coupling or dephasing may also appear. 

The one-dimensional counterpart of the MZM is the chiral Majorana edge mode (CMEM), which emerges as the topological edge state in two-dimensional $p$-wave superconductors\cite{read2000paired}. 
Such one-dimensional CMEMs usually possess linear dispersion near the Fermi energy and hence are more favorable in electrical transport. It has been theoretically proved that the propagation of the CMEMs could also lead to the similar non-Abelian transformations as those in the braiding process of the ``zero-dimensional'' MZMs\cite{lian2018topological,1DMajorana}. Instead of the delicate manipulations required for braiding MZM, the braiding operation can be naturally accomplished through the swift propagation of CMEMs, which efficiently enhance the speed and stability of the quantum gates\cite{lian2018topological}. 
Moreover, the possible signals of the CMEMs have been reported in the hybrid system of quantum anomalous Hall insulator (QAHI) and superconductor\cite{he2017chiral}. Based on the above reasons, the CMEMs now has also served as a possible platform for non-Abelian braiding.

Nonetheless, the experimental signals\cite{he2017chiral} for CMEMs are still controversial since some non-Majorana mechanism\cite{kayyalha2020absence} cannot be excluded yet. Such controversy in Majorana-based braiding inspire ones to explore the possible non-Abelian statistics in non-Majorana systems where the superconductivity is absent. Remarkably, topological insulators supporting Dirac edge states share a similar Hamiltonian with the topological superconductors supporting Majorana edge states, although they are described in two different Hamiltonian basis. Such an analogy implies the similarities in both their topological and transport properties\cite{lee2007edge,wu2020double}. Actually, the ``zero-dimensional'' Dirac fermionic mode as the topological end states of one-dimensional topological insulators, which is regarded as the ``Dirac counterpart'' of the MZM, also obeys the non-Abelian braiding statistics\cite{NucPhysB859261, wu2020double, Dirac0D} as one might have expected. 

In contrast to the Majorana case in which more experimental evidences for both the MZM\cite{PhysRevLett.109.267002, ruby2015end, PhysRevB.97.165302, PhysRevB.97.214502, vuik2019reproducing} and the CMEM\cite{kayyalha2020absence} are still highly requested, the  topological chiral Dirac edge modes (CDEMs) have been repeatedly verified in a number of experimental platforms including integer quantum Hall (IQH)\cite{QHI,novoselov2007roomQHI,tsukazaki2007oxideQHI} insulators and QAHIs\cite{QAHI,kou2014QAHI,bestwick2015QAHI,checkelsky2014QAHI} in a solid way. Since Dirac fermionic modes have been demonstrated to obey non-Abelian braiding statistics\cite{NucPhysB859261, wu2020double, Dirac0D}, in analogy to the idea ``promoting'' MZMs into CMEMs, we can naturally wonder whether we can extend topological Dirac fermionic modes from ``zero-dimension'' to one-dimension. Such a one-dimensional extension is exactly the CDEM discussed above. By noticing again the similarity between the topological insulators and the topological superconductors, one can also expect that the CDEMs will exhibit non-Abelian properties as CMEMs do during their scattering and propagation process. 

In this paper, we investigate the CDEM-based braiding both analytically and numerically. We also propose a viable experimental scheme for its realization. In Sec. \ref{sec_analytical}, we analytically calculate the scattering matrix between CDEMs and discuss the resonant condition for realizing the non-Abelian braiding operation; in Sec. \ref{sec_experiment}, based on our experimental proposal, we discover that the resonant condition is accompanied with the conductance peak, which could be satisfied by tuning the back gate voltages; in Sec. \ref{sec_numerical}, we conduct a numerical simulation of our experimental device and investigate the effects of different parameters, which corroborates our analytical conclusion. in Sec. \ref{sec_discussions}, we make a brief discussion comparing the braiding properties of the CDEMs and the MZMs; and finally in Sec. \ref{sec_conclusion}, we present a short conclusion.


\section{Analytical study for the scattering between chiral Dirac edge modes} \label{sec_analytical}
\begin{figure}[t]
    \centering
    \includegraphics[width=7.3cm]{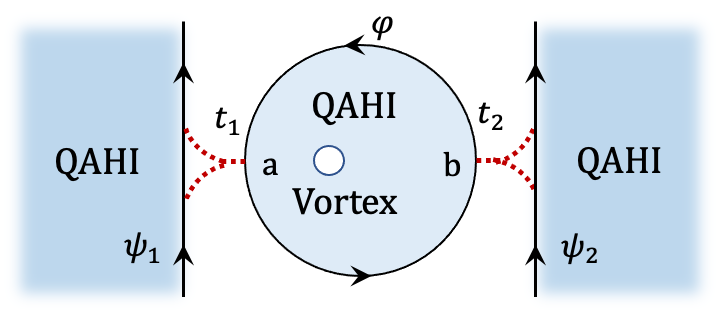}
    \caption{A QAHI island with magnetic vortices penetrated is placed in the center, which possesses a branch of CDEM $\varphi$ at its outer boundary. Another two CDEMs are coupled to the QAHI islands through weak tunneling at points $a$ and $b$ with tunneling amplitudes $t_1$ and $t_2$, respectively.}
    \label{fig:setup}
\end{figure}

Following a similar path of the CMEM-based quantum gates, we intend to implement the braiding operation through the scattering and propagation of CDEMs. Therefore, we begin with the scattering process of CDEMs coupled in the manner shown in Fig. \ref{fig:setup}, which bears a resemblance to the Andreev reflection induced by the CMEM\cite{Andreev}. A QAHI island with some magnetic vortices penetrated serves as the scattering center here, where the edge state of this QAHI island is a CDEM. Another two CDEMs as the edge states of another two QAHI bulks are coupled to the scattering center at points $a$ and $b$ with tunneling amplitudes $t_1$ and $t_2$, respectively. The total Hamiltonian of the system is

\begin{eqnarray} \label{tot}
H_{\mathrm{tot}} = -i\hbar v_f\int_{-\infty}^{\infty}[\psi^\dagger_1(x)\partial_x \psi_1(x)+\psi^\dagger_2(x)\partial_x \psi_2(x)]\mathrm{d}x \quad & \nonumber \\
+i\hbar v \int_0^L \varphi^\dagger(x) \partial_x \varphi(x)\mathrm{d}x -it_1 [\varphi^\dagger(a)\psi_1(0)+ \varphi(a)\psi_1^\dagger(0)]  \nonumber & \\
-it_2 [\varphi^\dagger(b)\psi_2(0)+\varphi(b) \psi_2^\dagger(0)] \quad \quad &
\end{eqnarray}

\noindent where $\varphi$ stands for the CDEM in the QAHI island and $\psi_1, \psi_2$ represents the CDEMs of QAHI coupled to the center.

Denoting the incoming and outgoing scattering states of the electrons by $\psi_{1/2\omega}(0-),\psi_{1/2\omega}(0+)$, respectively ($\omega$ stands for the electron's energy), the scattering matrix can be written as
\begin{equation}
{
\left( \begin{array}{c}
\psi_{1\omega}(0+)  \\
\psi_{2\omega}(0+)  
\end{array} 
\right )} =S{
\left( \begin{array}{c}
\psi_{1\omega}(0-)  \\
\psi_{2\omega}(0-)  
\end{array} 
\right )}
\end{equation}
Using the anticommutation relation and the equation of motion for the field operators, the explicit form of the $S$-matrix\cite{datta1997electronic} can be derived as:

\begin{widetext}
\begin{eqnarray}
S=\frac{1}{Z}{
\left( \begin{array}{ccc}
i \sin(\frac{\alpha+\beta}{2})(1-\tilde{t}_1^2\tilde{t}_2^2)-(\tilde{t}_1^2-\tilde{t}_2^2) \cos(\frac{\alpha+\beta}{2}) & -2e^{i(\beta-\alpha)/2}\tilde{t}_1\tilde{t}_2 \\
-2e^{i(\alpha-\beta)/2}\tilde{t}_1\tilde{t}_2 & i \sin(\frac{\alpha+\beta}{2})(1-\tilde{t}_1^2\tilde{t}_2^2)-(\tilde{t}_2^2-\tilde{t}_1^2) \cos(\frac{\alpha+\beta}{2}) 
\end{array} 
\right )}
\label{Smatrix_general_form}
\end{eqnarray}
\end{widetext}

\noindent where $Z=i \sin(\frac{\alpha+\beta}{2})(1+\tilde{t}_1^2\tilde{t}_2^2)+(\tilde{t}_1^2+\tilde{t}_2^2) \cos(\frac{\alpha+\beta}{2})$. $e^{i\alpha}(e^{i\beta})$ is the phase factor acquired by a CDEM propagating from point $a(b)$ to point $b(a)$. $\tilde{t}_1=t_1/2\hbar\sqrt{v v_f}$, and $\tilde{t}_2=t_2/2\hbar\sqrt{v v_f}$ are dimensionless coupling amplitudes.

Then we denote $\alpha+\beta$ by $\theta$, which stands for the overall phase acquired by the CDEMs when they make a complete circle around the QAHI island. For a CDEM with energy $\omega$, we have $\theta=\omega L/v+\Phi+\pi$, where $v$ is the Fermi velocity of the QAHI island, $L$ stands for the perimeter of the QAHI island, $\Phi$ is the phase induced by the magnetic flux penetrating the QAHI island, and the last term comes from the Berry phase contribution of the spin\cite{Andreev}. 

In some special cases, the $S$-matrix [Eq. (\ref{Smatrix_general_form})] can reduce to a simple form which is significant for the realization of non-Abelian braiding. We first consider a symmetric case in which $\tilde{t}_1=\tilde{t}_2$, with the overall phase $\theta=2m\pi$ ($m \in \mathbb{Z}$), then the $S$-matrix becomes $S=\left(\begin{matrix} 0 & (-1)^{m+1}e^{i(\beta-\alpha)/2} \\ (-1)^{m+1}e^{i(\alpha-\beta)/2} & 0 \\ \end{matrix}\right)$, which implies a resonant exchange with $\psi_1 \to (-1)^{m+1}e^{i(\alpha-\beta)/2}\psi_2$ and $\psi_2 \to (-1)^{m+1}e^{i(\beta-\alpha)/2}\psi_1$. Furthermore, for a single-coupled case where $\tilde{t}_1=0$, and the overall phase $\theta=2m\pi (m \in \mathbb{Z})$ is the same as the previous case, it is evident that $S=\left(\begin{matrix} 1 & 0 \\ 0 & -1 \\ \end{matrix}\right)$, indicating that the phase shift induced by the scattering is $\pi$. Remarkably, if we conduct the above two manipulations successively, the CDEMs $\psi_1$ and $\psi_2$ will be scattered into $(-1)^{m+1}e^{i(\alpha-\beta)/2}\psi_2$ and $(-1)^{m}e^{i(\beta-\alpha)/2}\psi_1$, respectively. After applying a gauge transformation that $\tilde{\psi}_1=e^{-i\alpha/2}\psi_1$, $\tilde{\psi}_2=e^{-i\beta/2}\psi_2$, finally we realize a non-Abelian braiding between two branches of CDEMs as $\tilde{\psi}_1 \to (-1)^{m+1} \tilde{\psi}_2$ and $\tilde{\psi}_2 \to (-1)^m \tilde{\psi}_1$.

According to the calculation above, in order to realize the non-Abelian braiding operation, it is essential for our system to satisfy the resonant condition $\theta=\alpha+\beta=2m\pi$ ($m \in \mathbb{Z}$). Since $\theta=\omega L/v+\Phi+\pi$ as mentioned above, such a resonant condition can be achieved by tuning the chemical potential of the QAHI island (the scattering center) to choose the specific electron energy that satisfies the requirement, or to change the number of the magnetic flux vortices penetrating through the QAHI island. 

So the crucial point is how to distinguish whether the resonant condition is satisfied or not experimentally, that is, how to manipulate the experimental parameters to conduct the braiding operation in a correct manner.


\section{Proposal for experimental scheme}  \label{sec_experiment}
\begin{figure}
    \centering
    \subfigure[]{
        \begin{minipage}[t]{5.5cm}
            \centering
            \includegraphics[width=5.8cm]{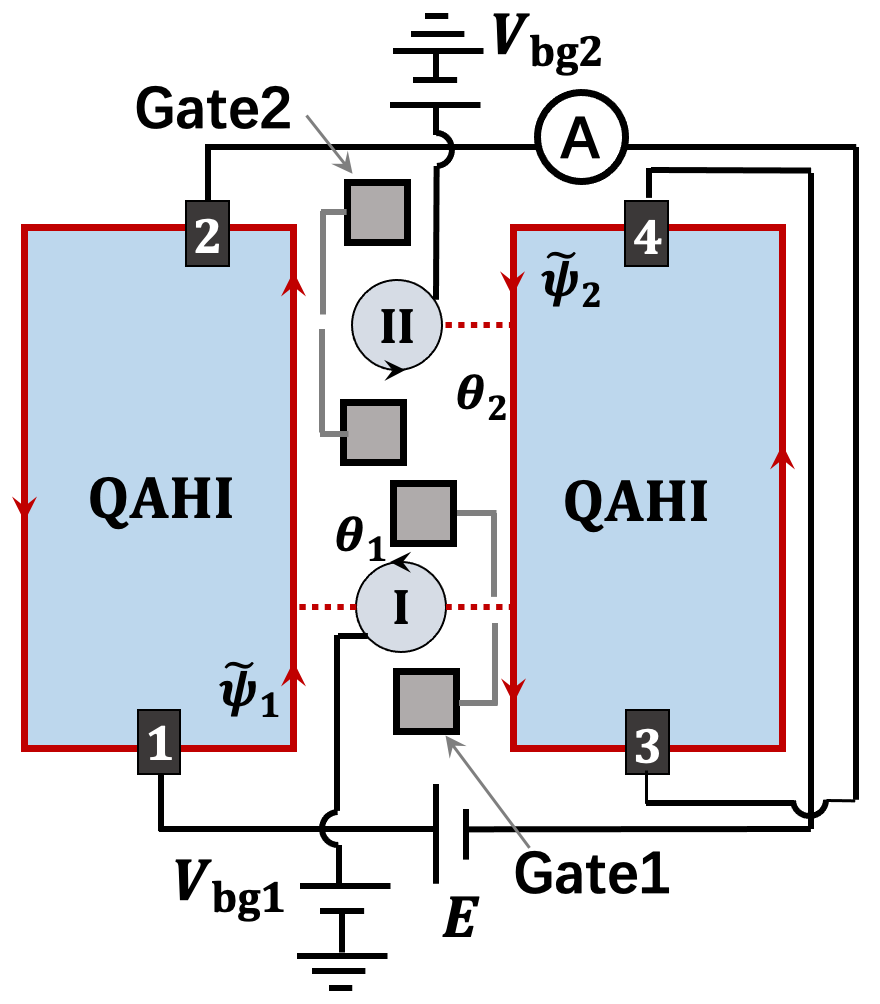}
        \end{minipage}
        \label{fig:device}
    }
    \subfigure[]{
        \begin{minipage}[t]{5.5cm}
            \centering
            \includegraphics[width=5.4cm]{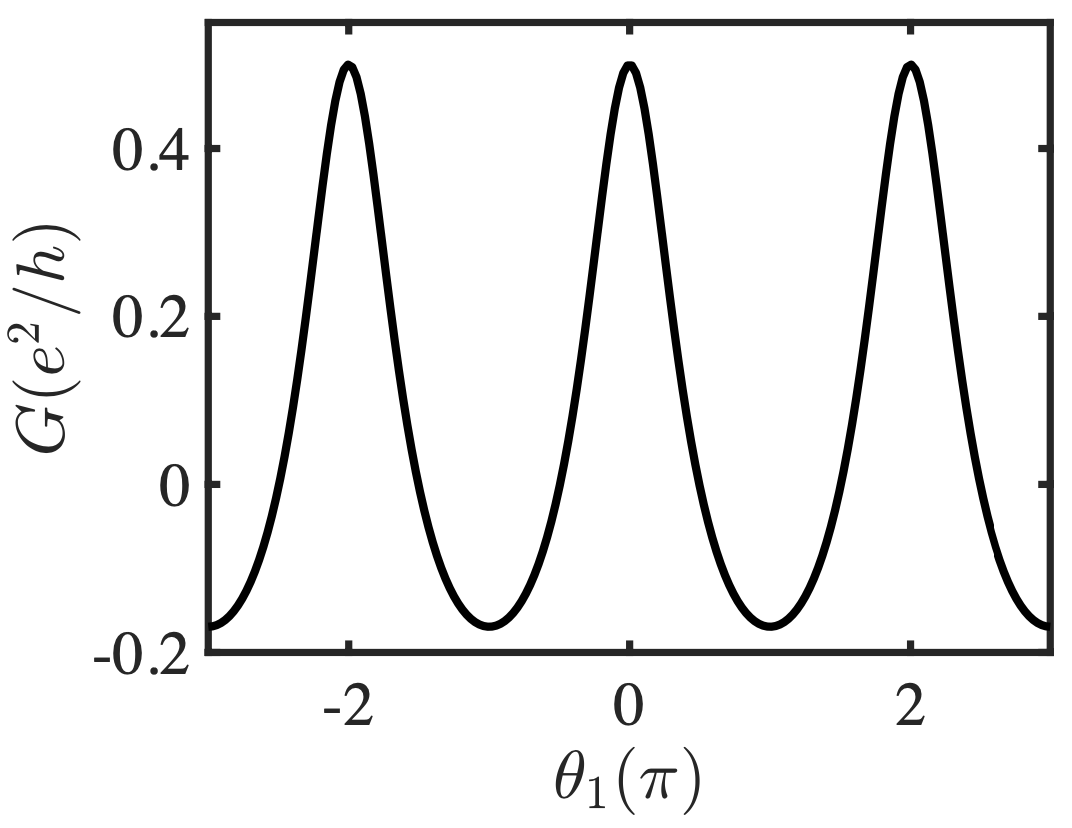}
        \end{minipage}
        \label{fig:I2}
    }
    \caption{(a) An experimental device based on QAHI to realize braiding operations on two CDEM branches $\tilde{\psi}_1,\tilde{\psi}_2$. Two QAHI islands are denoted by I and II, the corresponding phases accumulated by circulating these two QAHI islands are denoted by $\theta_1$ and $\theta_2$, respectively. Back gate $V_{bg1}$ ($V_{bg2}$) is attached to QAHI island I (II) for tuning $\theta_1$ ($\theta_2$). 
    Two pairs of split gates (denoted as ``Gate 1'' and ``Gate 2'') are deposited to manipulate the coupling between $\tilde{\psi}_1$ and QAHI islands II, and the coupling between $\tilde{\psi}_2$ and QAHI islands I, respectively. We define the state when gate voltage is applied and the coupling is cut off as ``Gate off'', otherwise as ``Gate on''. (b) The tunneling conductance $G=I_2/\Delta V$ as a function of $\theta_1$.}
    \label{setup}
\end{figure}

Based on the discussion above, we propose a four-lead device to observe the non-Abelian braiding of CDEMs experimentally. As shown in Fig. \ref{fig:device}, the device consists of four blocks of QAHI with Chern number $\mathcal{C}=1$, which correspond to one CDEM branch in each QAHI block. Two QAHI islands (denoted by I, II) with magnetic vortices penetrated serve as two scattering centers, and another two QAHI regions coupled to them provide two branches of CDEMs $\tilde{\psi}_1,\tilde{\psi}_2$. A bias $\Delta V$ is applied between lead 1 and lead 4, while lead 2 and lead 3 are shorted. These two CDEMs are scattered and manipulated during its propagation along the QAHI edges, and their transport behaviors can be detected by measuring the conductance. Two pairs of split gates are deposited at the coupling sites to form the gate-defined confinement\cite{GateAnneal,bid2009shot}. In such a way, the tunneling between the scattering centers and $\tilde{\psi}_1,\tilde{\psi}_2$ could be manipulated by tuning the corresponding split gate voltages. 

Noticeably, we only take QAHIs here as an example, all the QAHI regions in Fig. \ref{fig:device} can be replaced by other materials supporting CDEMs, e.g., gallium arsenide heterostructure supporting IQH effect. In addition, the device in Fig. \ref{fig:device} could be experimentally realized by depositing a top gate with specific shape above a whole QAHI sample. By depleting the electrons underlying such a top gate, the whole QAH sample can be separated into four blocks as we desired. \cite{weaktunnel}

\begin{figure*}
    \centering
    \subfigure[]{
        \begin{minipage}[t]{6.5cm}
            \centering
            \includegraphics[height=5cm]{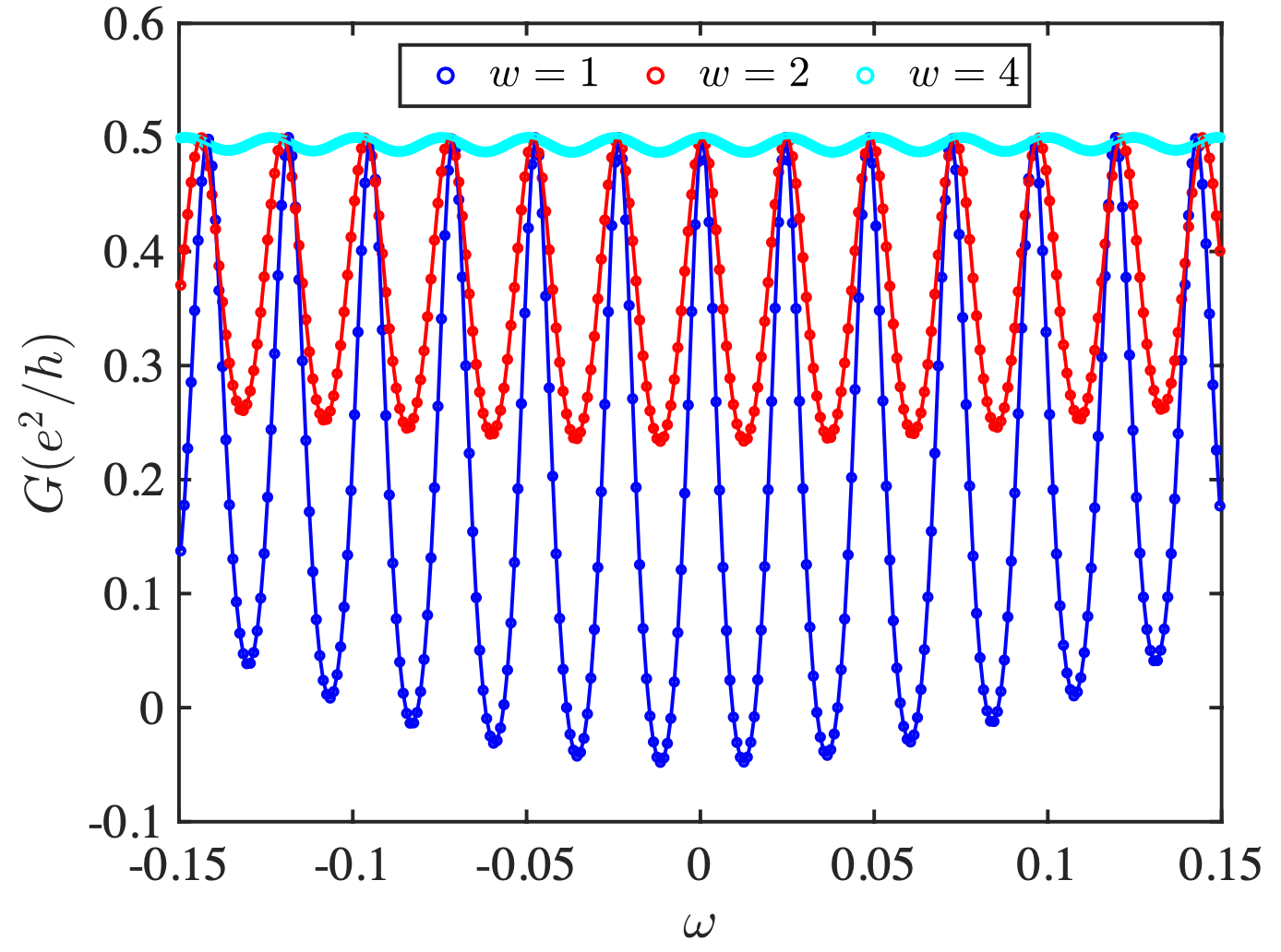}
        \end{minipage}
        \label{fig:width}
    }
    \subfigure[]{
        \begin{minipage}[t]{6.5cm}
            \centering
            \includegraphics[height=5cm]{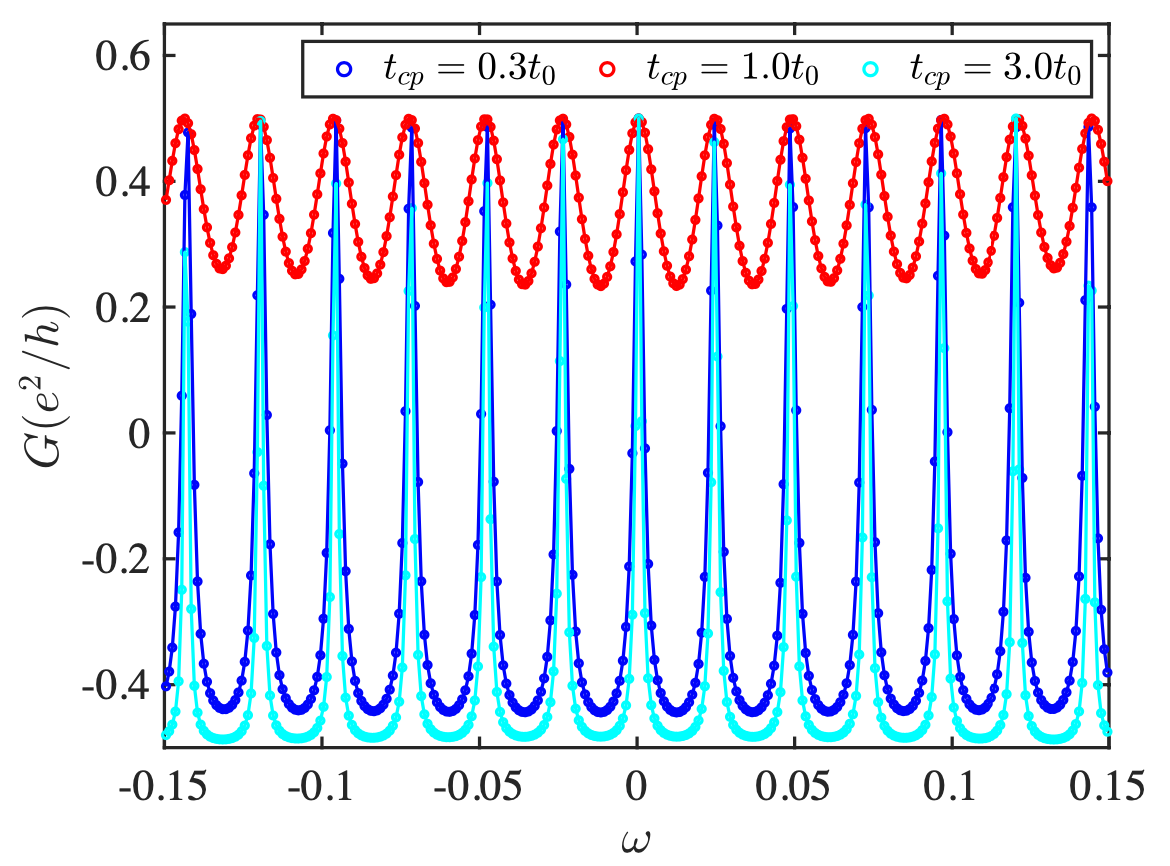}
        \end{minipage}
        \label{fig:strength}
    }
    \subfigure[]{
        \begin{minipage}[t]{6.5cm}
            \centering
            \includegraphics[height=5cm]{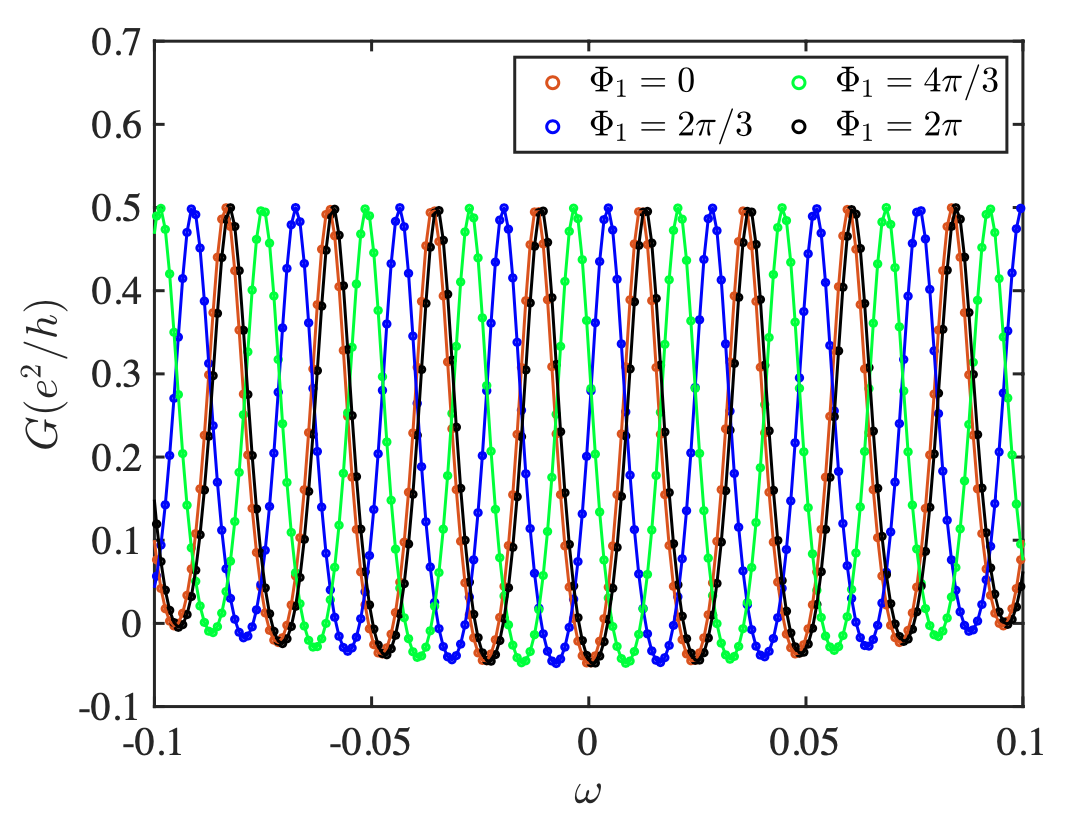}
        \end{minipage}
        \label{fig:flux}
    }
    \subfigure[]{
        \begin{minipage}[t]{6.5cm}
            \centering
            \includegraphics[height=5cm]{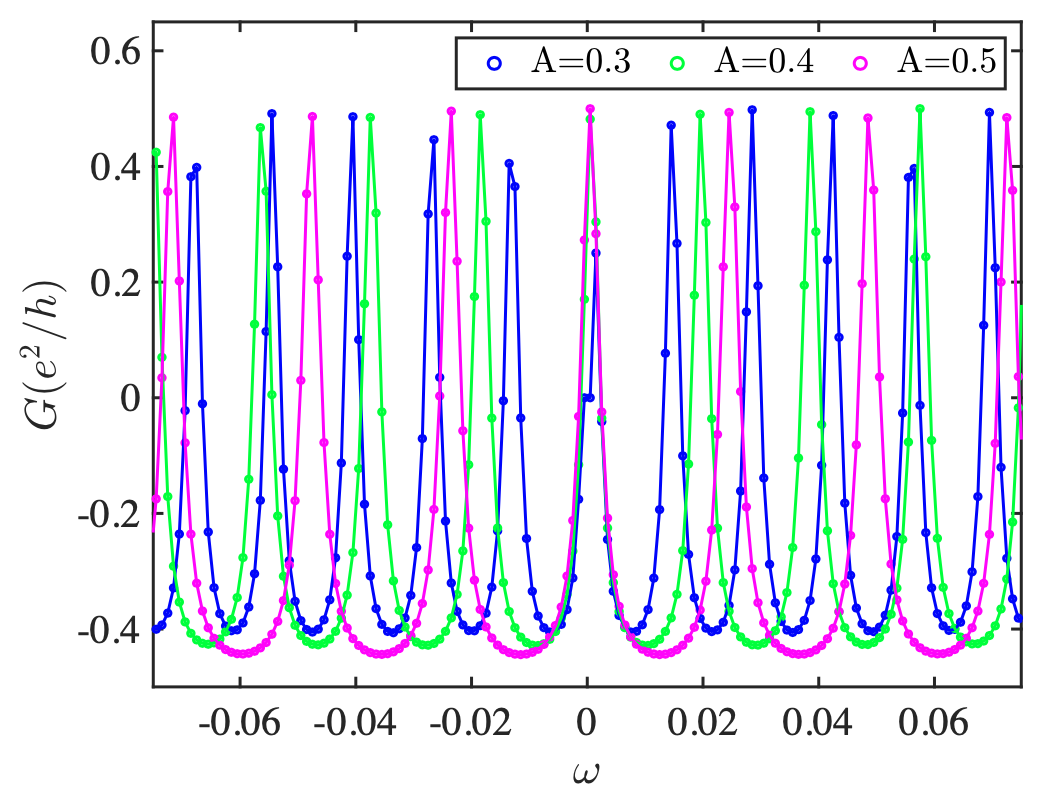}
        \end{minipage}
        \label{fig:para}
    }
    \caption{The numerical results of the tunneling conductance $G$ as a function of the electron's energy $\omega$, with respect to different connection width $w$, coupling strength $t_{\mathrm{cp}}$, the phase $\Phi_1$ induced by the magnetic flux in the QAHI island I, and the energy-band parameter $A$. (a) Tunneling conductance $G$ versus $\omega$ with respect to different $w$, with $t_{\mathrm{cp}}=1.0t_0, \Phi_1=\pi, A=0.5$. ($t_0=-\frac{i}{2}A\sigma_y+B\sigma_z$ with $B$=0.5, and the same below.) (b) Tunneling conductance $G$ versus $\omega$ with respect to different $t_{\mathrm{cp}}$, with $w=2, \Phi_1=\pi, A=0.5$. (c) Tunneling conductance $G$ versus $\omega$ with respect to different $\Phi_1$, with $w=1, t_{\mathrm{cp}}=1.0t_0, A=0.5$. (d) Tunneling conductance $G$ versus $\omega$ with respect to different $A$, with $w=2, t_{\mathrm{cp}}=1.0t_0, \Phi_1=\pi$.}
    \label{numerical}
\end{figure*}

To observe the effect of the phase $\theta_1$ (the phase acquired by the CDEMs when circulating the QAHI island I), we first consider the case with Gate 2 off and Gate 1 on so that the tunneling amplitude between $\tilde{\psi}_1$ and QAHI islands II (controlled by Gate 2) is pinched off, while the tunneling between $\tilde{\psi}_2$ and QAHI islands I (manipulated by Gate 1) is allowed [see Fig. \ref{fig:device}]. Under this condition, the scattering of island II just applies an additional phase to $\tilde{\psi}_2$, thus the tunneling conductance is independent of $\theta_2$.
The measured currents of lead $n$ can be calculated using the Landauer-Buttiker formula\cite{Landauer-Buttiker}: $I_n=\frac{e^2}{h}\sum_{m}T_{nm}(V_n-V_m)$, where tunneling coefficients $T_{nm}$ can be deduced from the scattering amplitude calculated above. 

Therefore the tunneling conductance $G$, which is defined as $I_2/\Delta V$, can be expressed as a function of $\theta_1$ as:
\begin{equation} \label{current}
\begin{aligned}
G&=\frac{I_2}{\Delta V}=\frac{e^2}{2h}(T_{32}-T_{31})\\
&=\frac{e^2}{2h}\frac{4\tilde{t}_1^2 \tilde{t}_2^2-(1-\tilde{t}_1^2\tilde{t}_2^2)^2 \sin^2\frac{\theta_1}{2}-(\tilde{t}_1^2-\tilde{t}_2^2)^2 \cos^2\frac{\theta_1}{2}}{(1+\tilde{t}_1^2\tilde{t}_2^2)^2 \sin^2\frac{\theta_1}{2}+(\tilde{t}_1^2+\tilde{t}_2^2)^2 \cos^2\frac{\theta_1}{2}}
\end{aligned}
\end{equation}

\noindent Fig. \ref{fig:I2} represents the oscillation behavior of the tunneling conductance $G$ as a function of $\theta_1$, in the condition that $\tilde{t}_1^4=\tilde{t}_2^4=\tilde{t}^4=0.1$.

From Fig. \ref{fig:I2}, one can find that when $\theta_1$ approaches $2m\pi$ ($m\in Z$), the tunneling conductance $G$ shows a peak of one half of $\frac{e^2}{h}$, that is, the resonant condition $\theta_1=2m\pi$ is satisfied when the conductance reaches its maximum. Such a resonant condition can be achieved by tuning the electron's energy $\omega$ through the back gate $V_{\mathrm{bg1}}$. 

The tuning of $\theta_2$ can be accomplished in the same fashion as $\theta_1$. We can first switch on Gate 2 and switch off Gate 1, thus the dependence of $G$ on $\theta_2$ is the same as that on $\theta_1$ in the previous case; then the resonant condition of $\theta_2$ can be achieved by tuning $V_{\mathrm{bg2}}$, and finally close Gate 2 and open Gate 1. 
Consequently, in our experimental scheme, the resonant condition can be reflected by the conductance peaks which are experimental observables. By switching on and off the gate-defined confinement and then modulating the back gate voltages, we can tune the phases $\theta_1$ and $\theta_2$ for both two QAHI islands to satisfy the resonant condition. Therefore we have proposed a set of viable procedures for the realization of the CDEM-based non-Abelian braiding operation.


\section{Numerical Evidence} \label{sec_numerical}

To verify the validity of our braiding proposal, we utilize a discretized lattice version of the Bernevig-Hughes-Zhang (BHZ) model\cite{bernevig2006quantum} to perform numerical simulation. The effective lattice Hamiltonian in each QAHI block is 

\begin{equation}
\begin{aligned}
H_{\mathrm{eff}}&= \sum_{\vec{k}} [A\sigma_x \sin k_x+A\sigma_y \sin k_y \\
&+(\Delta -4B \sin^2 \frac{k_x}{2}-4B \sin^2\frac{k_y}{2})\sigma_z ]|\vec{k}\rangle \langle \vec{k}| \\
\end{aligned}
\end{equation}

\noindent In addition to the four QAHI blocks, the coupling of the CDEMs $\tilde{\psi}_1,\tilde{\psi}_2$ and the two QAHI islands can be included by adding corresponding hopping terms connecting the corresponding QAHI block boundaries with connection width $w$ and hopping strength $t_{\mathrm{cp}}$. To obviate the effects of bulk states, the energy $\omega$ of the lattice Green function in our numerical investigation lies within the bulk gap. 

Our numerical results are shown in Fig. \ref{numerical}, the tunneling conductance $G$ exhibits a periodic oscillation as expected, and its maximum coincides with our theoretical prediction. The  amplitude and the full width at half maximum (FWHM) of each peak varies with the connection width $w$ [Fig. \ref{fig:width}] as well as the hopping strength $t_{\mathrm{cp}}$ [Fig. \ref{fig:strength}], which implies that the effective coupling strength $t_1, t_2$ in the analytic model [see Eq. (\ref{tot})] depends on both $w$ and $t_{\mathrm{cp}}$. With the increase of the connection width $w$, the CDEM  will tunnel into the QAHI island from different lattice sites with different phases and then a self-interference is presented. Hence the FWHM decreases with the increase of $w$, as shown in Fig. \ref{fig:width}.

However, an unexpected damping of the oscillation amplitude accompanies with the oscillating behavior, where the damping rate differs with respect to different parameter choices. A reasonable explanation is that the Fourier transform of the coupling strength from the form in the tight-binding model $t_{\mathrm{cp}}$ to the form in the analytic model $t_1, t_2$ depends on the energy $\omega$. Hence the effective coupling strength in the lattice model varies with the energy $\omega$.

Nonetheless, the phase $\theta_1$ remains unchanged with the variance of the coupling strength transformation, so that both the oscillation period and the phase shift are independent of the energy $\omega$, the coupling width $w$ and the coupling strength $t_{\mathrm{cp}}$, which is consistent with the simulation results shown in Fig. \ref{fig:width}, \ref{fig:strength}. 

\begin{figure}[t]
    \centering
    \subfigure[]{
        \begin{minipage}[t]{4.2cm}
            \centering
            \includegraphics[height=3.5cm]{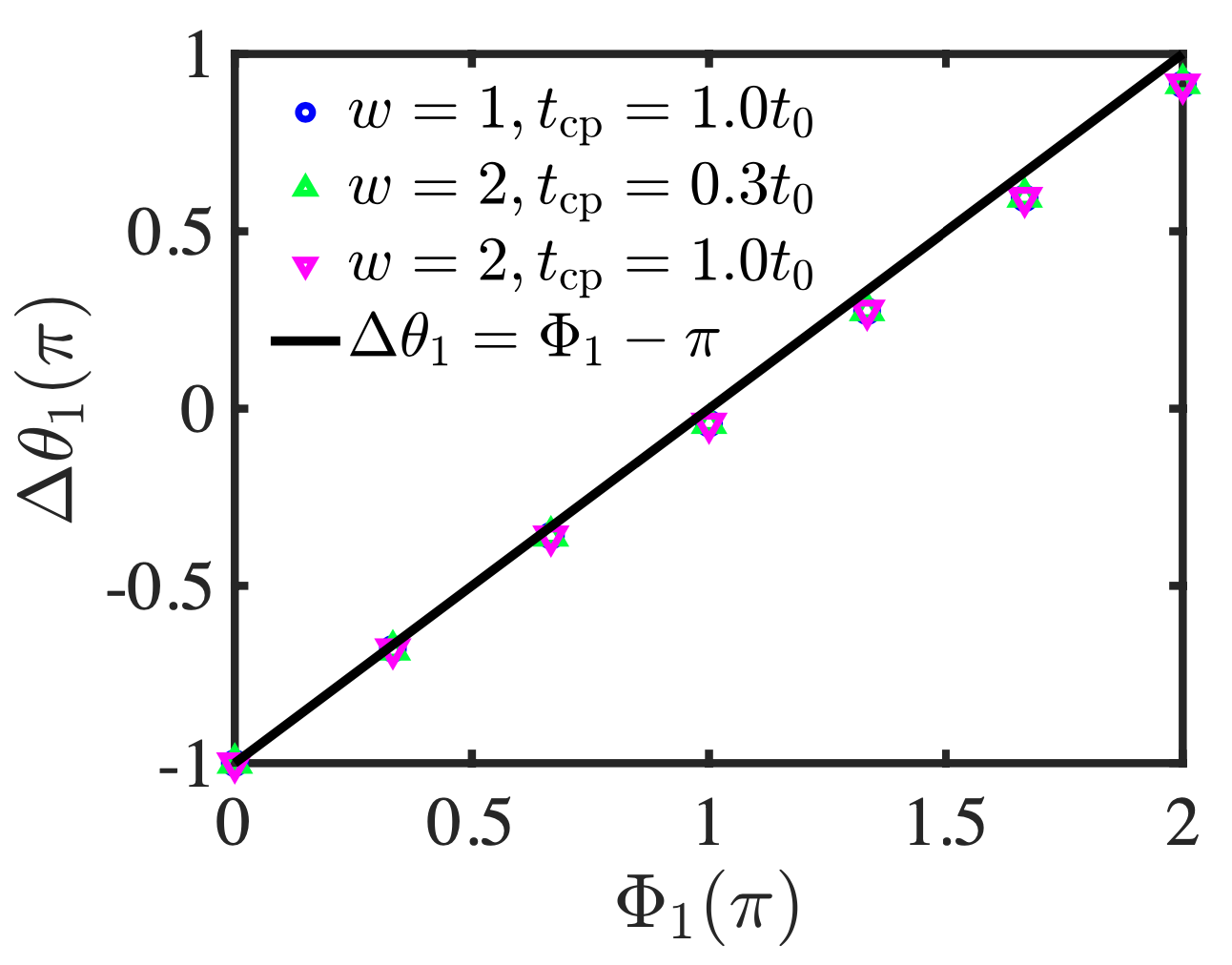}
        \end{minipage}
        \label{fig:phase}
    }
    \subfigure[]{
        \begin{minipage}[t]{3.8cm}
            \centering
            \includegraphics[height=3.5cm]{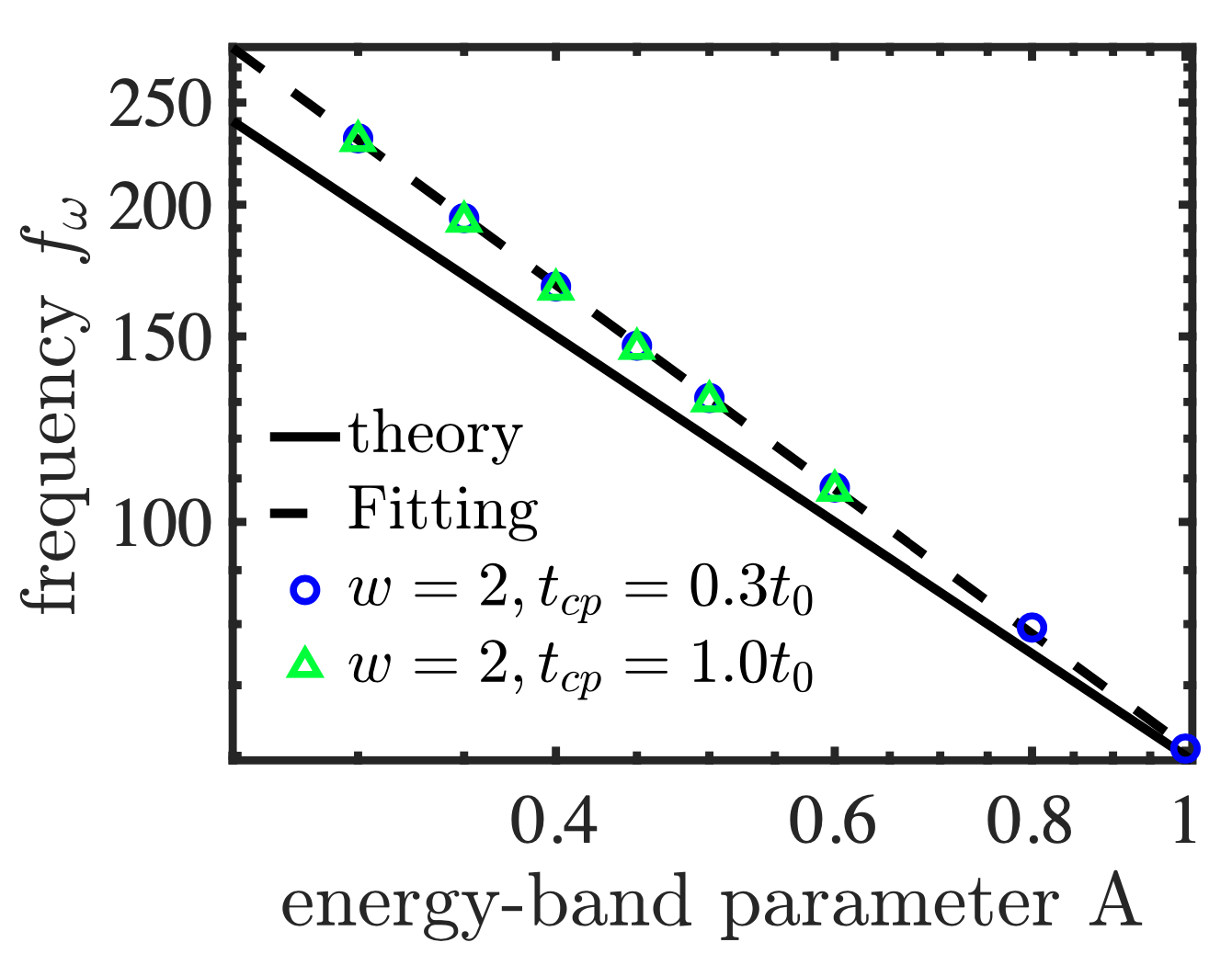}
        \end{minipage}
        \label{fig:period}
    }
    \caption{(a) The phase shift with respect to the phase induced by the flux inside the QAHI island I $\Phi_1$ under different $w$ and $t_{\mathrm{cp}}$. The consistency between the numerical simulation and the theoretical formula confirms that the numerical results come from the physics of edge states. (b) The oscillation frequency of the current with respect to energy-band parameter $A$ under different coupling conditions. }
    \label{finalres}
\end{figure}

As shown in Fig. \ref{fig:flux}, the phase shift of the conductance curve can be induced by the magnetic flux $\Phi_1$ penetrating through the QAHI island I. The phase shift is defined as $\Delta \theta_1 = \theta_1(\Phi_1)-\theta_1(\Phi_1=\pi)$. If the CDEMs are well localized at the edge so that all the magnetic flux penetrated is encircled, then the phase shift will follow such a compact relation: $\Delta \theta_1=\Phi_1-\pi$. We calculate the phase shifts with respect to different magnetic fluxes and different coupling parameters [Fig. \ref{fig:phase}], and all of them are in good agreement with the formula $\Delta \theta_1 = \Phi_1-\pi$, corroborating that the tunneling conductance obtained in the numerical simulation is contributed from the CDEMs which circulate all the magnetic flux penetrated through the QAHI island. 


Furthermore, in the low-energy limit, the oscillation frequency $f_\omega$ with respect to energy $\omega$ equals to $\frac{L}{2A}$. Thus by tuning the energy-band parameter $A$, we can modulate the oscillation period of the tunneling conductance $G$ as the numerical results shown in Fig. \ref{fig:para}. As shown in Fig. \ref{fig:period}, our simulation results exhibit a strong linear relation in log-log plot with a slope that is approximately equal to $1$. Such results corroborate our analytical formula. Furthermore, based on the fitting  results shown in Fig. \ref{fig:period}, we can obtain an effective perimeter $L_{\mathrm{eff}}$ of the QAHI island, which roughly equals to the perimeter of our simulation model $L$ as $L_{\mathrm{eff}} \gtrsim L$, confirming the contribution from edge physics. The slight deviation between $L_{\mathrm{eff}}$ and $L$ may contributes to the tunneling points near which the CDEMs may make a detour so that the effective perimeter becomes longer. 


Now we conclude that the numerical results based on the BHZ model are in good agreement with the analytical ones, providing a more solid demonstration for our experimental scheme.


\section{Discussions}
\label{sec_discussions} 

In the previous sections, we have investigated the CDEM-based non-Abelian braiding both analytically and numerically, and proposed viable device for further experiments. Compared with other braiding schemes, the CDEM-based braiding inherit the favorable transport properties of the CMEMs that the braiding operation can be accomplished via its natural edge propagation other than the external manipulation. 
Furthermore, the experimental platforms supporting CDEMs possess desirable robustness for their larger bulk gap\cite{wu2014prediction,qi2016high,li2020high} compared with the Majorana system. The absence of the superconductivity in the CDEM-based braiding not only significantly reduce the complexity of the experimental setup, but also exclude other possible states \cite{PhysRevB.97.165302, PhysRevB.97.214502, vuik2019reproducing} that may obstruct the non-Abelian braiding. 

In addition, we have demonstrated the non-Abelian properties of CDEMs, which bears resemblance to the braiding properties of the renowned MZMs. Whereas, it is noteworthy that the corresponding basis of non-Abelian braiding operation for CDEMs \cite{NucPhysB859261, Dirac0D} differ from those for MZMs. The basis for CDEM-based braiding are all single-particle operators, thus the braiding operation preserves the particle number. In comparison, the MZM-based braiding preserves the electron parity, which allows particles created or annihilated in pairs\cite{Ivanov}.

Furthermore, in our experimental proposal, the braiding effect is detected by observing the conductance peak due to resonant tunneling. Although resonant tunneling could also be exhibited by a localized state bound in a quantum dot (QD)\cite{1DMajorana}, the QD will only induce a single peak rather than a series of equally-spaced conductance peaks with the same height shown in our model. Hence the conductance signal characteristics possessed by the CDEMs in our proposal is distinguishable from those come from the other trivial mechanisms.  



\section{Conclusion}
\label{sec_conclusion} 

In conclusion, we have demonstrated the non-Abelian braiding operation can be conducted utilizing CDEMs. Through two successive scattering process between two CDEMs, the overall operation recovers the standard non-Abelian form under a gauge transformation. Additionally, the effect of the braiding operation can be experimentally detected by observing the conductance peaks in QAHI junctions, which has been demonstrated by simulations. 
Compared with the MZM-based braiding proposal, our scheme overcomes two main obstacles in topological quantum computing: the speed of braiding operation and the robustness of edge modes. By utilizing the high velocity of the CDEMs and the robust experimental platforms supporting CDEMs, our braiding proposal enables us to manipulate CDEMs with desirable speed and stability, thus may sheds some light on the topological quantum computation.


\section{Acknowledgements} 

We thank for the fruitful discussion with Hua Jiang. This work is financially supported by the Strategic Priority Research Program of Chinese Academy of Sciences (Grant No. XDB28000000), the National Basic Research Program of China (Grants No. 2015CB921102), and the China Postdoctoral Science Foundation (Grant No. 2021M690233)

\bibliography{nonAbelian_CDM}

\end{document}